\documentclass[a4paper, amsfonts, amssymb, amsmath, reprint, showkeys, nofootinbib, twoside]{revtex4-1}
\usepackage[english]{babel}
\usepackage[utf8]{inputenc}
\usepackage[colorinlistoftodos, color=green!40, prependcaption]{todonotes}
\usepackage{amsthm}
\usepackage{mathtools}
\usepackage{physics}
\usepackage{xcolor}
\usepackage{graphicx}
\usepackage[left=23mm,right=13mm,top=35mm,columnsep=15pt]{geometry} 
\usepackage{adjustbox}
\usepackage{placeins}
\usepackage[T1]{fontenc}
\usepackage{lipsum}
\usepackage{csquotes}
\usepackage[pdftex, pdftitle={Article}, pdfauthor={Author}]{hyperref} 

\usepackage{amsmath}
\usepackage{soul}
\usepackage{booktabs}
\usepackage[english]{babel}

\begin{document}
\title{Cepstral Analysis for Baseline-Insensitive Absorption Spectroscopy Using Light Sources with Pronounced Intensity Variations}

\author{Christopher S. Goldenstein}
\email{csgoldenstein@purdue.edu}
\author{Garrett C. Mathews}%
\affiliation{School of Mechanical Engineering, Purdue University, 585 Purdue Mall, West Lafayette, IN 47907, USA}%

\author{Ryan K. Cole}
\author{Amanda S. Makowiecki}
\author{Gregory B. Rieker}
\affiliation{Precision Laser Diagnostics Laboratory, University of Colorado Boulder, 1111 Engineering Drive, Boulder, CO 80309, USA}%

\date{\today} 

\begin{abstract}
This manuscript presents a data-processing technique which improves the accuracy and precision of absorption-spectroscopy measurements by isolating the molecular absorbance signal from errors in the baseline light intensity ($I_o$) using cepstral analysis. Recently, cepstral analysis has been used with traditional absorption spectrometers to create a modified form of the time-domain molecular free-induction decay (m-FID) signal which can be analyzed independently from $I_o$. However, independent analysis of the molecular signature is not possible when the baseline intensity and molecular response do not separate well in the time domain, which is typical when using injection-current-tuned lasers (e.g., tunable diode and quantum cascade lasers) and other light sources with pronounced intensity tuning. In contrast, the method presented here is applicable to virtually all light sources since it determines gas properties by least-squares fitting a simulated m-FID signal (comprising an estimated $I_o$ and simulated absorbance spectrum) to the measured m-FID signal in the time domain. This method is insensitive to errors in the estimated $I_o$ which vary slowly with optical frequency and, therefore, decay rapidly in the time domain. The benefits provided by this method are demonstrated via scanned-wavelength direct-absorption-spectroscopy measurements acquired with a distributed-feedback (DFB) quantum-cascade laser (QCL). The wavelength of a DFB QCL was scanned across CO's P(0,20) and P(1,14) absorption transitions at 1 kHz to measure the gas temperature and concentration of CO. Measurements were acquired in a gas cell and in a laminar ethylene-air diffusion flame at 1 atm. The measured spectra were processed using the new m-FID-based method and two traditional methods which rely on inferring (instead of rejecting) the baseline error within the spectral-fitting routine. The m-FID-based method demonstrated superior accuracy in all cases and a measurement precision that was $\approx$1.5 to 10 times smaller than that provided using traditional methods. 
\end{abstract}

\maketitle

\section{Introduction}

Laser-absorption spectroscopy (LAS) is a powerful and broadly applicable technique for providing quantitative measurements of gas conditions and species concentrations \cite{Allen1998,Werle1998,Lackner2007,Bolshov2015a, Goldenstein2017a}. While there are many variants of LAS, all methods ultimately rely on discerning how much of the incident light was absorbed by the test gas, typically as a function of wavelength. This is most obvious in the context of direct-absorption-spectroscopy (DAS) techniques which rely on converting a measurement of the transmitted light intensity ($I_t$) to the spectral absorbance ($\alpha$) of the test gas using Beer's Law (Eq. \ref{eq: Beers}). The incident (i.e., baseline) light intensity ($I_o$) must be well known to execute this conversion accurately and, ultimately, to provide an accurate measurement of gas properties.

While rarely discussed in the literature, accurate determination of $I_o$ frequently limits the accuracy and precision of LAS diagnostics. Time- or wavelength-dependent variations in $I_o$ due to etalon interference effects (e.g., produced by windows or other planar optical components) and/or variations in the laser's intensity can be difficult to account for with high accuracy, even in tame laboratory environments. For general context, many LAS applications such as characterization of thermochemical flame structure and combustion kinetics, atmospheric sensing, and high-fidelity characterization of absorption lineshapes often demand measurements with <2\% error. In the case of measuring a spectral absorbance of 0.05 (typical of near-infrared LAS applications), achieving a 2\% error in the spectral absorbance of the target species requires achieving an effective error in $I_o$ of only 0.1\%. This is especially challenging to achieve when characterizing harsh combustion environments where beamsteering, window fouling, mechanical vibration, and scattering off particulates frequently cause the transmitted light intensity to vary on the order of 1 to 10\% \cite{Ruesch2020} and, in extreme applications (e.g., coal gasifiers, explosive fireballs), by several orders of magnitude \cite{Sun2013a,Mathews2019,Lodes2019}.

To overcome this challenge, researchers have developed a variety of strategies to avoid or mitigate the impact of measurement errors induced by uncertainty or error in $I_o$. For example, wavelength-modulation spectroscopy (WMS) with various harmonic-normalization techniques (e.g., WMS-2\textit{f}/1\textit{f}, RAM-normalization) \cite{Cassidy1982,Rieker2009, Goldenstein2014,Upadhyay2018,Peng2020} can actively account for variations in $I_o$ induced by broadband transmission losses. These methods can be especially advantageous in circumstances where the absorbance spectra are spectrally broad compared to the wavelength-scan amplitude of the laser and non-resonant wavelengths cannot be reached (e.g., in gases at high-pressures) \cite{Goldenstein2017a}. For example, researchers have demonstrated that WMS-2\textit{f}/1\textit{f} is capable of providing high-fidelity measurements of gas properties in a variety of high-pressure combustion environments where narrowband lasers such as distributed-feedback (DFB) tunable diode lasers (TDLs) and quantum-cascade lasers (QCLs) cannot access a non-absorbing baseline or interrogate a sufficiently large portion of the spectrum to reliably infer the baseline via post-processing \cite{Rieker2007,Sun2013a,Goldenstein2015d,Lee2018a}. That being said, WMS techniques remain susceptible to error induced by background signals originating from, for example, a non-linear laser-intensity response or intensity modulation induced by etalon effects, which can be time-varying and non-trivial to account for (e.g., using background subtraction or accounting for background signals in the WMS model). Such errors ultimately stem from a lack of understanding of $I_o$ or the signal components comprising it \cite{Goldenstein2014}.

In the context of DAS, numerous strategies have been developed to mitigate errors induced by $I_o$ \cite{Ebert2000,Wagner2009,Teichert2003,Goldenstein2013e,Schulze2005,Blume2016,Emmert2018,Simms2015,Rieker2014,Kauppinen2011,MosierBoss1995,Kranendonk2007b,Cole2019a,Wang2020}. Perhaps the most widely used method utilizes a polynomial or spline to account for wavelength-dependent variations in $I_o$. This method has been widely utilized in both narrowband techniques (e.g., TDLAS) \cite{Ebert2000,Wagner2009,Teichert2003,Goldenstein2013e} and broadband techniques (e.g., using FTIR, frequency-combs, supercontinuum lasers) \cite{Schulze2005,Blume2016,Emmert2018,Simms2015,Rieker2014}, although the latter typically employs multiple polynomials in a piecewise-fitting approach. The polynomial(s) can be determined prior to least-squares fitting a simulated absorbance spectrum to the measured absorbance spectrum (e.g., by fitting to the non-absorbing regions of $I_t$) or they can be determined along with the best-fit absorbance spectrum simultaneously using the spectral-fitting routine. The latter approach is less susceptible to user bias and Simms et al. \cite{Simms2015} demonstrated it can reduce measurement error. In any case, these methods are susceptible to errors induced by coupling between the polynomial(s) and the simulated absorbance spectrum. This is particularly problematic when the spectroscopic model used for calculating the best-fit absorbance spectrum is heavily flawed, in which case the fitting routine may erroneously attribute errors in the absorbance model to errors in $I_o$ (accounted for with the polynomial). Alternatively, the fitting routine may simply converge on an inaccurate solution that simply leads to the smallest sum-of-squared error. Ultimately the modeled spectrum may match the measured spectrum very well, however the gas properties inferred from the best-fit spectrum could have large errors. 

Other methods for correcting errors in the baseline leverage differences in the spectral "shape" between the absorbance spectrum and $I_o$. For example, Kranendonk et al. \cite{Kranendonk2007b} analyzed the first derivative of the absorbance spectrum to desensitize the measurement to errors in $I_o$ that vary slowly with frequency compared to the absorbance spectrum. As a result, this method is best suited for cases where the absorbance spectra consist of discrete, spectrally narrow lines (e.g., from small molecules at low pressures). One disadvantage of this technique is that the signal must be smoothed after differentiation as this process can amplify noise. Another method, utilizes Fourier transforms and bandpass filtering to effectively separate the absorbance spectrum from the baseline intensity \cite{Schulze2005, Kauppinen2011,MosierBoss1995}. In this approach, first the Fourier transform of the measured transmission spectrum is calculated. If the spectrum consists of discrete absorption features, the absorption lines occur at specific frequencies (in the signal's power spectrum) which are then isolated using bandpass filters. After bandpass filtering the signal from the absorption lines, the inverse Fourier transform of this signal is calculated to yield a corrected spectrum which is less prone to baseline errors. That being said, the corrected spectrum must still be normalized to account for the baseline intensity prior to comparing with simulated absorbance spectra for determination of gas properties.


Most recently, Cole et al. \cite{Cole2019a} developed a technique which eliminates the need to account for the baseline in post-processing. This method works by converting the measured transmitted intensity spectrum ($I_t$) to a modified form of the molecular free-induction decay using cepstral analysis. The modified free-induction decay (m-FID) signal consists of two distinct components with an additive relationship: (1) the laser-intensity response (from $I_o$) and (2) the molecular-absorption response (from $\alpha$). In the time domain, these signals can separate from each other since the laser-intensity response can decay to zero more rapidly. The authors showed that this enables gas properties to be determined, without knowledge of $I_o$, by least-squares fitting a simulated molecular-absorption response signal (obtained from a simulated absorbance spectrum only) to the molecular-absorption response within the measured m-FID signal. This approach was demonstrated with broadband absorption measurements (synthetic and real) of species (e.g., CH$_4$, C$_2$H$_6$) with discrete and/or quasi-continuous absorbance spectra using a dual-frequency-comb spectrometer with complex frequency-dependent variations in $I_o$. That said, achieving baseline-free measurements with this technique is limited to cases where the laser-intensity response decays to zero in the time domain faster than the molecular-absorption response. 

The work presented here builds upon the m-FID-based approach developed by Cole et al. \cite{Cole2019a} in order to accommodate scenarios where the laser-intensity response and the molecular-absorption response decay on similar timescales and, therefore, do not fully separate in the time domain. In contrast to the method of Cole et al. \cite{Cole2019a}, this method relies on modeling the entire m-FID signal using an estimated $I_o$ (e.g., from baseline fitting) and a simulated absorbance spectrum, and least-squares fitting the simulated m-FID signal to the measured m-FID signal. As such, this method is not "baseline-free," however we demonstrate that this approach reduces measurement errors significantly by separating the molecular-absorption response from error in the estimated $I_o$. Most importantly, this approach is applicable to scenarios with large and rapid (with optical frequency) variations in $I_o$ such as are encountered in scanned-wavelength direct-absorption experiments conducted with injection-current-tuned semiconductor lasers (e.g., TDLs, QCLs). As such, this method enables the error-reducing benefits of m-FID-based analysis to be attained in a wider variety of LAS experiments. The remainder of this manuscript is devoted to describing the fundamentals and operating principles of this method, as well as to presenting the experimental validation of this technique and comparison with established data-processing methods.



\begin{figure*}[!t]
\centering
\includegraphics[width=18cm]{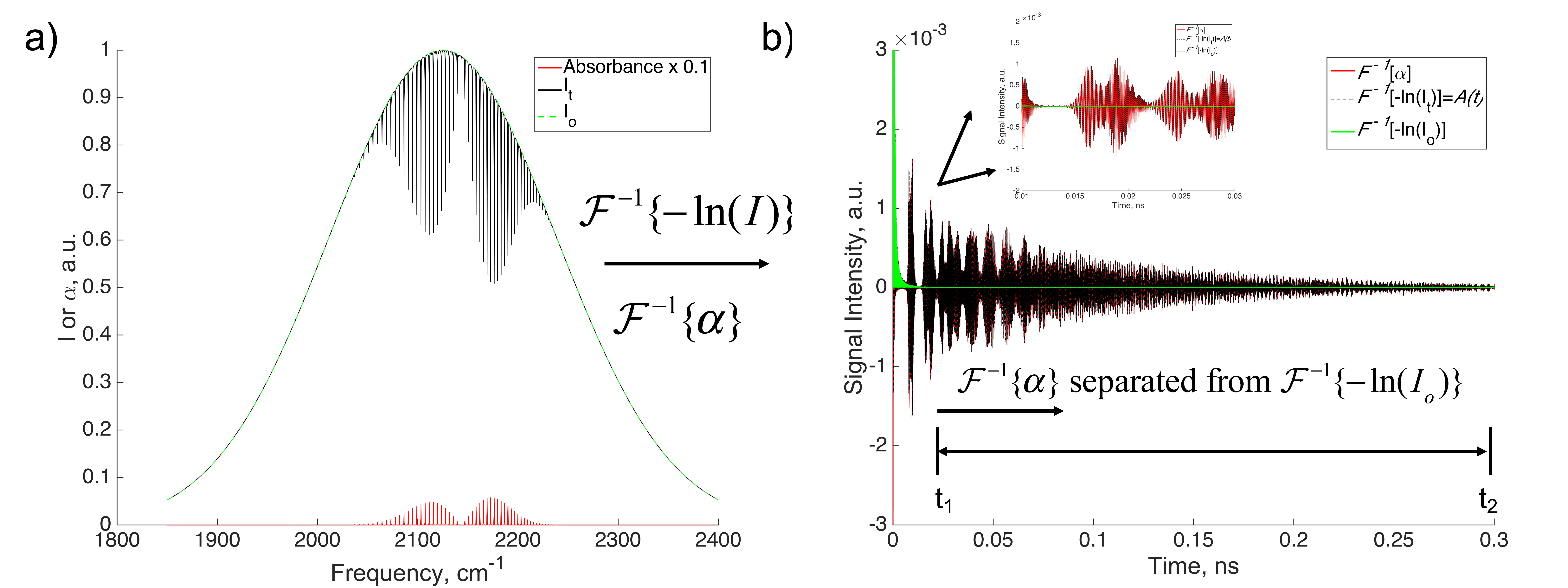}
\caption{(a) Intensity ($I_o$, $I_t$) and absorbance spectra for a simulated ultrafast laser-absorption measurement of CO spectra near 2150 cm$^{-1}$ using a transform-limited 55 fs pulse. (b) Zoom view of the beginning of the m-FID signal, laser-intensity response, and molecular-absorption response in the time domain which correspond to the spectra shown in the optical frequency domain in (a).}
\label{fig: UltrafastTandmFID}
\end{figure*}

\section{Fundamentals of Absorption Spectroscopy and m-FID}

This section describes the pertinent fundamentals of absorption spectroscopy and how the m-FID signal is related to $I_o$ and absorbance spectra.

\subsection{Absorption Spectroscopy}

In LAS, a monochromatic laser beam with incident intensity $I_o$ and frequency $\nu$ is directed through a gas sample and the transmitted light intensity $I_t$ is measured on a photodetector. Beer's Law, given by Eq. \ref{eq: Beers}, can be used to determine the spectral absorbance, $\alpha$, and for a uniform line-of-sight it is related to spectroscopic parameters and thermodynamic properties using Eq. \ref{eq: alpha}.

\begin{equation} 
I_t(\nu) =I_o(\nu)exp[-\alpha(\nu)]
\label{eq: Beers}
\end{equation}

\begin{equation} 
\alpha(\nu)=\sum_jS_j(T)P\chi_{Abs}\phi_j(\nu)L
\label{eq: alpha}
\end{equation}

\noindent Here, $S_j$ (cm$^{-2}$/atm) is the linestrength of transition $j$ at temperature $T$, $P$ (atm) is the pressure of the gas, $\chi_{Abs}$ is the mole fraction of the absorbing species, $\phi_j$ (cm) is the lineshape of transition $j$, and $L$ (cm) is the path length through the gas sample. Thermodynamic properties of the test gas (e.g., T, $\chi_{Abs}$) can be determined by comparing measured absorbance spectra to modeled absorbance spectra, for example, using a spectral-fitting routine such as those described in Section \ref{Sect: FitRoutine}.

\subsection{m-FID Signal}

\subsubsection{Calculating the M-FID Signal}

Recent work by Cole et al. \cite{Cole2019a} introduced the m-FID signal (which derives from the traditional time-domain free-induction-decay signal through cepstral analysis), $A(t)$, which is related to $I_t$, $I_o$, and $\alpha$ according to Eq. \ref{eq: Av} and \ref{eq: At}. 

\begin{equation} 
A(\nu) = -ln(I_t) = \alpha(\nu) - ln(I_o(\nu))
\label{eq: Av}
\end{equation}

\begin{equation} 
A(t) = \mathcal{F}^{-1}[A(\nu)] = \mathcal{F}^{-1}[\alpha(\nu)] + \mathcal{F}^{-1}[-ln(I_o(\nu))]
\label{eq: At}
\end{equation}

\noindent Here, $\mathcal{F}^{-1}$ represents the inverse Fourier Transform of a given quantity, $A(t)$ is the m-FID signal (also known as the Cepstrum of $I_t(t)$) \cite{Rosenblatt1963,Oppenheim2004}, $\mathcal{F}^{-1}[-ln(I_o(\nu))]$ is the laser-intensity response, and $\mathcal{F}^{-1}[\alpha(\nu)]$ is the molecular-absorption response. Eq. \ref{eq: Av} comes from taking $-ln$ of Eq. \ref{eq: Beers} (i.e., Beers Law). Eq. \ref{eq: At} illustrates that the m-FID signal can be found from taking the inverse Fourier Transform of $A(\nu)$ and that, in the time domain, there is an additive relationship between the molecular-absorption response and the laser-intensity response due to the logarithmic operation in the frequency domain. 

\subsubsection{M-FID Using an Ultrafast Pulse}
The physical meaning and behavior of the m-FID signal and its components are best understood by considering an experiment where a single ultrafast transform-limited pulse is used to measure the absorbance spectrum of a molecule (e.g., similar to as described in Tancin et al. \cite{Tancin2020}). That being said, it is important to note that the m-FID signal can be calculated from any measurement of a transmitted light intensity spectrum. To elucidate the principles governing the m-FID signal, this section will discuss a simulated experiment where a transform-limited pulse with a full-width at half-maximum of 55 fs in the time domain and 267 cm$^{-1}$ in the frequency domain is used to measure the absorbance spectrum of CO's fundamental vibration band near 2150 cm$^{-1}$ at a temperature and pressure of 300 K and 1 atm, respectively. Figure \ref{fig: UltrafastTandmFID}a illustrates simulated intensity and absorbance spectra corresponding to this simulated experiment and Figure \ref{fig: UltrafastTandmFID}b illustrates the corresponding m-FID signal in the time domain, which is composed of the laser-intensity response and the molecular absorption response.  In this case, the laser-intensity response (shown in green) is largest at time zero, decays rapidly on the timescale of the pulse FWHM, and reaches near zero (1\% of its initial intensity) within $\approx$1 ps. In comparison, the molecular-absorption response (shown in red) is also largest at time zero and rapidly decays to near zero immediately following the laser pulse. However, the magnitude of the signal quickly recovers and then periodically oscillates before permanently decaying to zero on the timescale of $\approx$ 1 ns. 

\begin{figure}[!b]
\centering
\includegraphics[width=8.5cm]{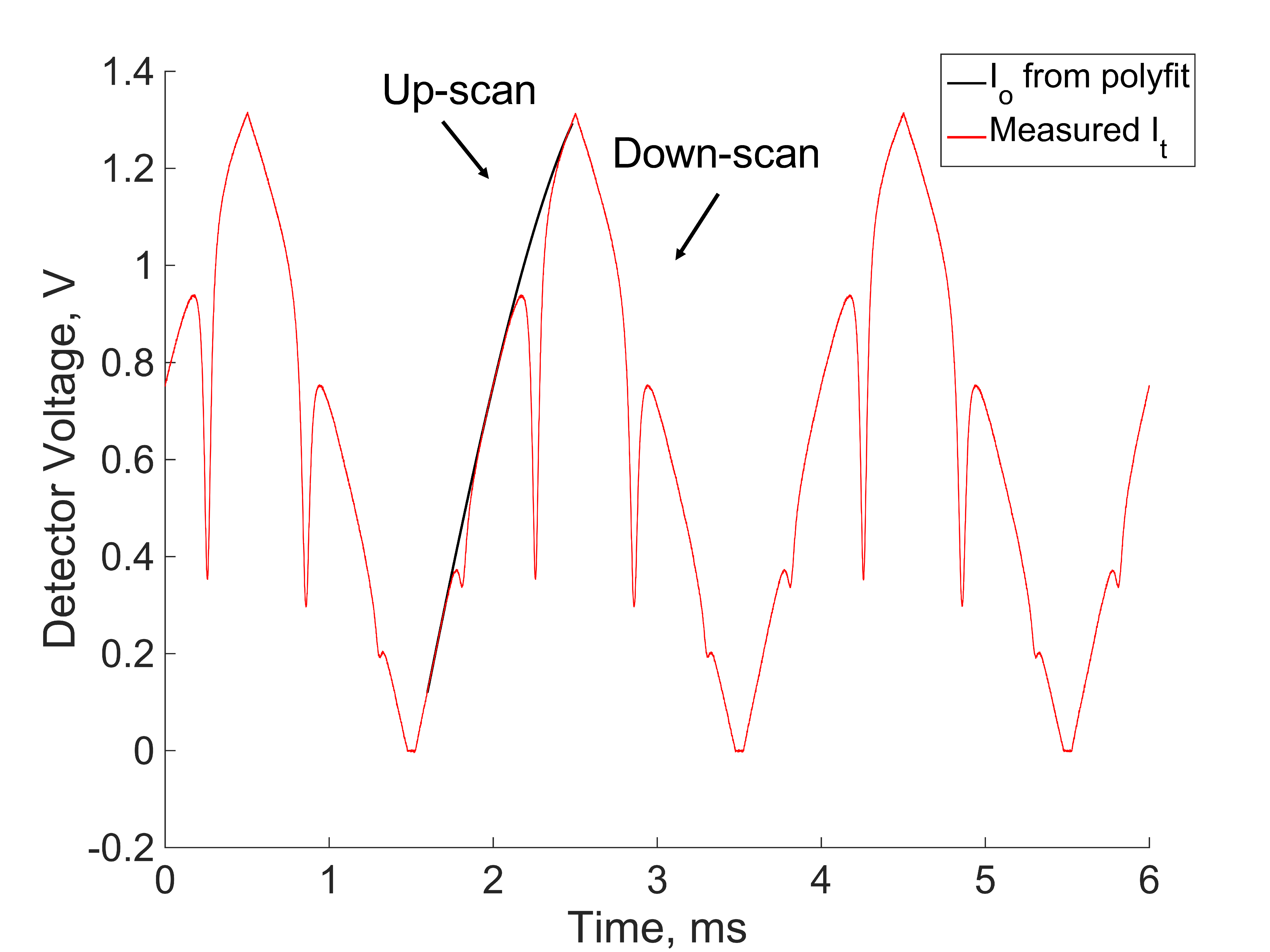}
\caption{Raw detector signal acquired in a scanned-DA experiment with a DFB QCL scanning across CO's P(0,20) and P(1,14) transitions at 500 Hz. The spectra were acquired in an ethylene-air flame at 1 atm.}
\label{fig: MultiScan}
\end{figure}

\begin{figure*}[!t]
\centering
\includegraphics[width=18cm]{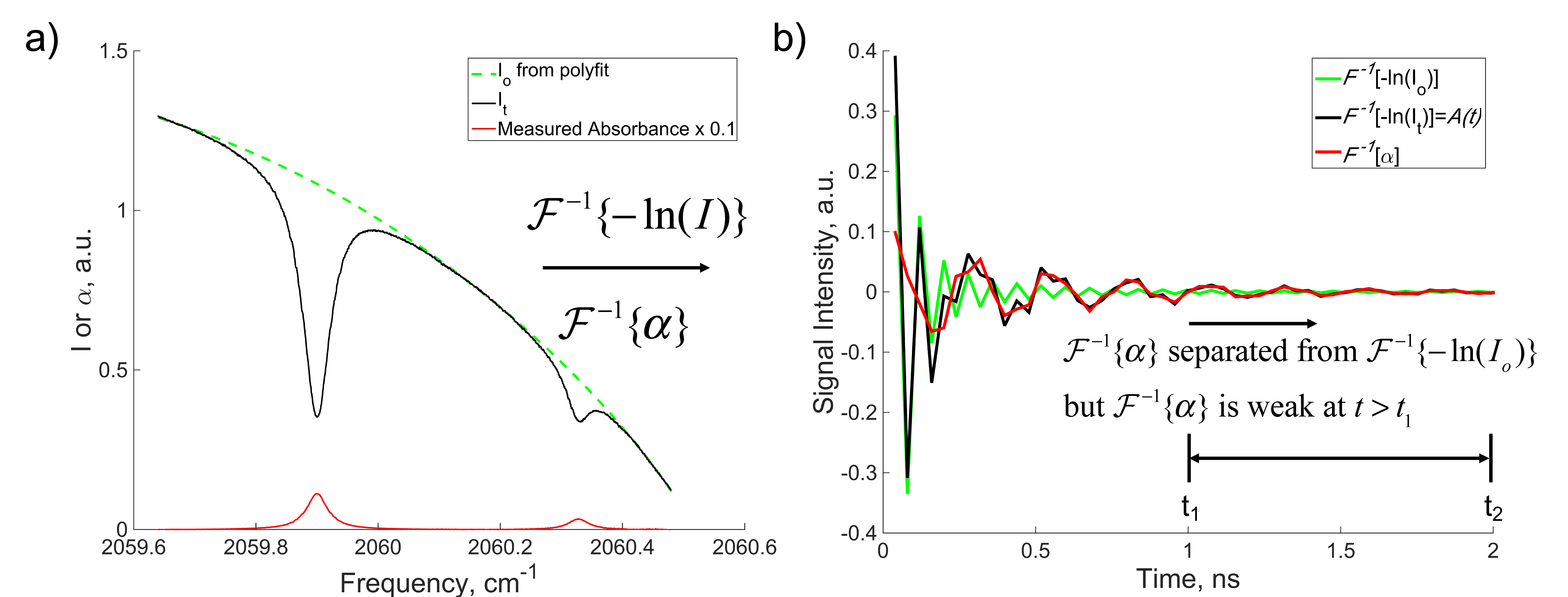}
\caption{(a) Measured $I_t$ for a single-scan across CO's P(0,20) and P(1,14) transitions, $I_o$ determined from baseline fitting, and corresponding absorbance spectrum. (b) Zoom view of the beginning of the m-FID signal, laser-intensity response, and molecular-absorption response in the time domain which correspond to the spectra shown in (a). The m-FID signal agrees well with the molecular-absorption response signal at times > 1 ns where the contribution from the laser-intensity response has decayed to near zero. The measurements were acquired in an ethylene-air flame at 1619 K, 1 atm, and with 11.1\% CO by mole.}
\label{fig: TandmFID}
\end{figure*}

This behavior can be understood by recognizing the following. In this simulated experiment, the broadband ultrashort pulse would near-instantaneously pump CO molecules into a wide range of quantized rovibrational states and the excited molecules would initially rotate in phase with each other and radiate coherently to emit a forward burst of radiation (i.e., the free-induction decay or m-FID signal) into the same mode as the laser pulse. Following the laser pulse, the molecules would rapidly rotate out of phase with each other (temporarily destroying the m-FID signal) due to state-specific differences in the rotational energy/frequency of the excited molecules. However, a short time later the molecules would then periodically re-phase with each other leading to additional forward bursts of coherent radiation (i.e., additional m-FID signal) \cite{Brewer1972,Picque2019,Coddington2010a,Coddington2010,Felkert1992}. Collisions and radiative decay would then permanently destroy the molecular coherence and ensuing m-FID signal that was created by the ultrashort pulse. The characteristic decay time of the m-FID signal due to collisional broadening only can be estimated using $\tau_{c,decay} \approx 1/\pi\Delta\bar{\nu_c}$ assuming an instantaneous excitation pulse of light where $\Delta\bar{\nu_c}$ [$s^{-1}$] is the average (across transitions) collisional-broadening (i.e., Lorentzian) full-width at half-maximum. This follows from relations put forth to model the free-induction decay signal \cite{Cole2019a}. In this simulated experiment, $\tau_{c,decay}$ = 0.108 ns which agrees reasonably well with the decay of the m-FID signal envelope shown in Figure \ref{fig: UltrafastTandmFID}b.


\subsubsection{M-FID Using an Injection-Current-Tuned Laser}

In practice, LAS experiments are often performed using injection-current-tuned lasers, for example, DFB TDLs and QCLs \cite{Goldenstein2017a, Werle1998}. In this case, injection-current scanning is performed to scan the frequency of the laser light and this also leads to pronounced intensity tuning. For example, Figure \ref{fig: MultiScan} shows the raw detector signal for a scanned-wavelength direct-absorption (scanned-DA) experiment performed using a DFB QCL which was scanned across CO's P(0,20) and P(1,14) absorption transitions near 2059.9 cm$^{-1}$. In this case, it is clear that the laser intensity varies rapidly with optical frequency on a scale that is comparable to that of the absorption lineshapes. In the context of the m-FID signal and its components, this translates into the laser-intensity response and molecular-absorption response decaying on a similar timescale.

Figure \ref{fig: TandmFID}a shows an example of a single-scan measurement of $I_t$, $I_o$ (inferred from fitting a polynomial to the non-absorbing regions of $I_t$), and the corresponding absorbance spectrum of CO near 2059.9 cm$^{-1}$ which were extracted from the data shown in Figure \ref{fig: MultiScan}. For comparison, Figure \ref{fig: TandmFID}b shows the beginning of the m-FID signal, laser-intensity response, and the molecular-absorption response in the time domain which correspond to the measured spectra shown in Figure \ref{fig: TandmFID}a. In this case, the laser-intensity response and molecular-absorption response decay on a similar timescale and are not well separated until $t_1 \approx$ 1 ns. In theory, the methods of Cole et al. \cite{Cole2019a} could be used to provide baseline-free measurements of gas properties by least-squares fitting simulations of the molecular-absorption response to the measured m-FID signal occurring between $t_1$=1 ns and $t_2$=2 ns; however far too much of the molecular-absorption response has decayed to zero by $t_1$= 1 ns for this method to yield an accurate measurement. In fact, this was attempted and the fitting routine failed to converge on a solution which motivated the development of the new technique described in Section \ref{Sect: FitRoutine}.




\section{Experimental Setup}

\begin{figure}[!t]
\centering
\includegraphics[width=8.5cm]{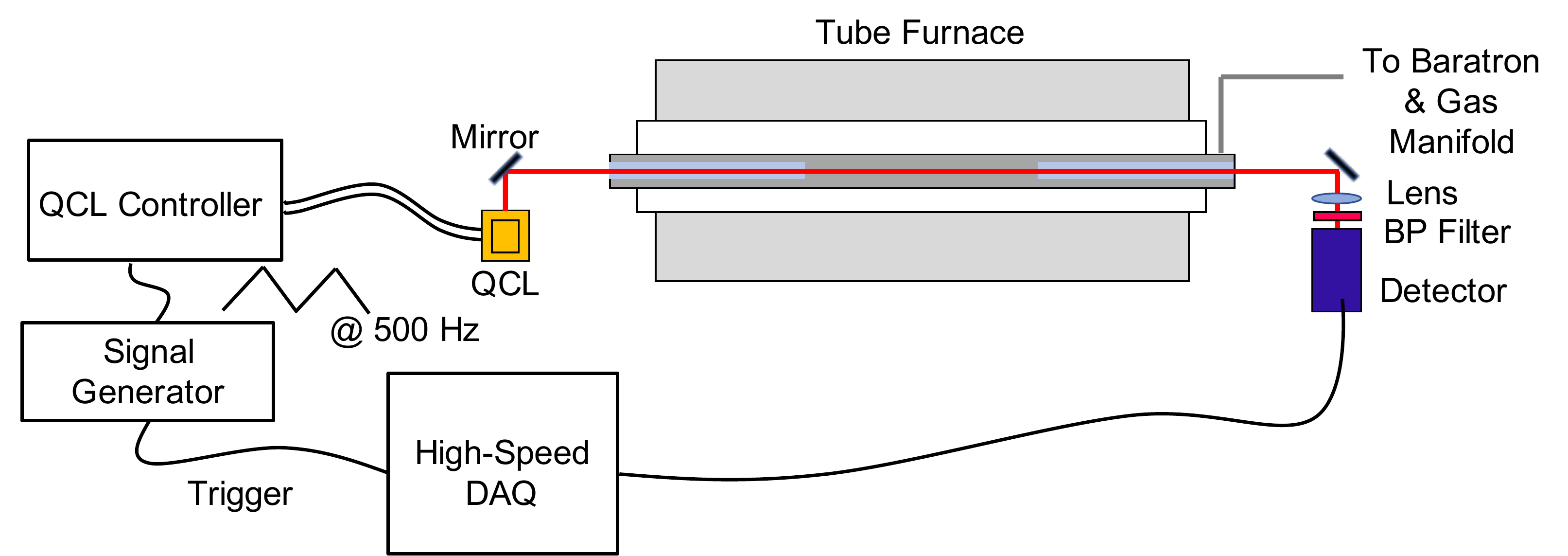}
\caption{Schematic of the experimental setup used to acquire scanned-DA measurements of gas temperature and CO mole fraction at 1 kHz in a heated static-gas cell.}
\label{fig: Setup}
\end{figure}


Figure \ref{fig: Setup} shows a schematic of the experimental setup used for gas-cell measurements. The wavelength of a distributed-feedback quantum-cascade laser (Alpes Lasers) was scanned across CO's P(0,20) and P(1,14) absorption transitions near 2060 cm$^{-1}$ to determine the gas temperature and concentration of CO. Several researchers have recently used these absorption transitions to provide high-quality measurements in a variety of combustion applications \cite{Spearrin2014c,Spearrin2014,Goldenstein2015d,Tancin2018a, Tancin2019a} and additional details regarding their suitability for high-temperature combustion gases are provided by Spearrin et al. \cite{Spearrin2014c}. The QCL produced a collimated laser beam ($\approx$1.5 mm diameter) with a maximum optical power of 50 mW. Wavelength scanning was performed through injection-current scanning by applying a 500 Hz, 0 to 600 mV triangle wave to the QCL current controller (Arroyo 6310). The wavelength scanning was characterized with a solid germanium etalon with a free-spectral range of 0.0165 cm$^{-1}$.  

The laser beam was directed through a static-gas cell located within a high-uniformity tube furnace. The gas cell is thoroughly described in \cite{Schwarm2019}, it employs sapphire or CaF$_2$ (used here) rods to ensure that the laser light propagates through a thermally uniform ($\pm$ 5 K) test section which is 9.4 cm long. Three type K thermocouples are mechanically fastened to the outer body of the gas cell along the test section to determine the gas temperature. Experiments were conducted using a professionally prepared (Airgas) mixture of 2\% CO by mole in N$_2$ at 1 atm and temperatures of 827 and 1034 K. The laser light exiting the gas cell was focused onto a photovoltaic MCT (mercury cadmium telluride) detector (Vigo Systems, PVI-5-1x1-TO8-BaF2) using an anti-reflection coated, plano-convex, CaF$_2$ lens (25.4 mm diameter, 30 mm focal length). The photodetector has a 3dB bandwidth of 10 MHz and it is sensitive to wavelengths from approximately 3 to 6 $\mu$m. The photodetector’s voltage signal was recorded using a 12-bit data-acquisition (DAQ) card (GaGe CSE123G2) with a bandwidth of 500 MHz and a sampling rate of 3 GS/s. Onboard averaging of the detector signal was performed to reach a final sampling rate of 1.875 MS/s and an effective bit depth of 16 bits for improved signal-to-noise ratio. A bandpass filter (Spectrogon) centered near 2060 cm$^{-1}$ with a FWHM of 40 cm$^{-1}$ was used to attenuate emission from the furnace and a thin ($\approx$2 mm) sheet of polycarbonate was used to attenuate the laser power and prevent detector saturation.

Measurements were also acquired in ethylene-air diffusion flames produced using a honeycomb burner with a square cross section. Fuel (C$_2$H$_4$) was passed through the core of the burner (0.5” wide cross section) and an air curtain (1” outer cross section) was used to stabilize the flame. The flow rates of air and fuel were manipulated to achieve a stable laminar flame. The laser beam was directed through the flame approximately 1 cm above the burner surface where the flame thickness (estimated from images of visible flame emission) was approximately 1.25 cm.

\section{Least-Squares Fitting to the m-FID Signal} \label{Sect: FitRoutine}

\subsection{Procedure}

This section describes our approach to determining gas properties from measured m-FID signals in circumstances where the laser-intensity response and molecular-absorption response do not separate quickly in the time domain. This is especially relevant to scenarios where the laser's intensity varies with optical frequency with a similar magnitude and spectral shape compared to the absorbance spectrum. This method builds on the fitting routine put forth by Cole et al. \cite{Cole2019a} by introducing one critical modification, specifically, an estimate for $I_o(\nu)$. In this method a simulated m-FID signal is generated using (1) an estimated $I_o(\nu)$ and (2) a simulated absorbance spectrum and this simulated m-FID signal is least-squares fit to the measured m-FID signal. The introduction of an estimated $I_o(\nu)$ allows the fitting routine to access more of the molecular-absorption response, which is particularly important if the laser-intensity and molecular-absorption responses are similar.  We will show that this approach is immune to baseline errors that vary slowly with frequency, and thus does not require a perfect estimate for the baseline. 
The remainder of this section is devoted to describing and demonstrating the fitting routine in detail.

\begin{figure}[!t]
\centering
\includegraphics[width=9cm]{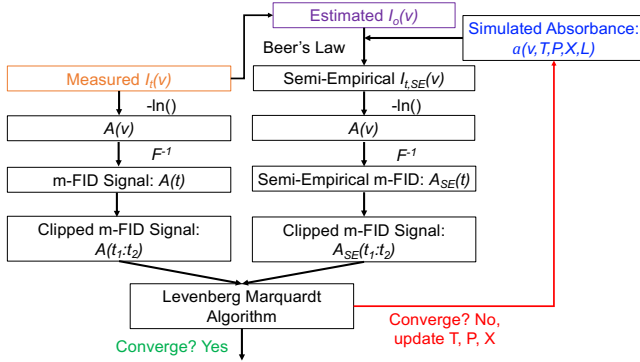}
\caption{Flowchart illustrating principles of the least-squares fitting routine used to determine gas properties from measured m-FID signals.}
\label{fig: FlowChart}
\end{figure}

Figure \ref{fig: FlowChart} illustrates a flow chart for the fitting routine used to determine gas properties from measured m-FID signals. Prior to calculating the measured m-FID signal, any background emission (e.g., from flame gas) must be subtracted from the measured detector signal to properly determine $I_t(\nu)$ as is traditionally the case. In addition, the measured spectrum of $I_t(\nu)$ must be re-sampled onto a frequency axis with uniform spacing (e.g., using interpolation). This is required in experiments performed with, for example, DFB QCLs or TDLs since their optical frequency does not vary exactly linearly with injection current, particularly at high scan rates or when using a large scan amplitude. Next, the m-FID signal corresponding to a measured spectrum of $I_t(\nu)$ must be calculated according to $\mathcal{F}^{-1}[-ln(I_t(\nu))]$ (see Eq. \ref{eq: At}). The inverse Fourier Transform should be calculated such that the m-FID signal is a purely real signal. This can be done using Python's function irfft or equivalent. 

The simulated m-FID signal should be calculated as follows. First, an estimate for the baseline light intensity $I_o(\nu)$ must be obtained. Here this was done by least-squares fitting a 3rd-order polynomial to the non-absorbing regions of the measured $I_t$ (see Figure \ref{fig: MultiScan}). Alternatively, a background measurement of $I_o(\nu)$ (e.g., in the absence of absorbing gas) could be used to determine an estimate for $I_o(\nu)$. Next, the absorbance spectrum must be calculated at gas conditions set by the free parameters. Here, the HITEMP2010 database \cite{Rothman2010} and a spectroscopic model similar to that described in \cite{Goldenstein2017} were used to simulate the absorbance spectrum of CO at the wavelengths of interest. Next, a simulated, semi-empirical spectrum of the transmitted light intensity $I_{t,SE}(\nu)$ was calculated using Eq. \ref{eq: Beers} with the estimated $I_o(\nu)$ and simulated $\alpha(\nu)$. A simulated, semi-empirical m-FID signal ($A_{SE}(t)$) was then calculated from $\mathcal{F}^{-1}[-ln(I_{t,SE}(\nu))]$.

\begin{figure*}[!t]
\centering
\includegraphics[width=18cm]{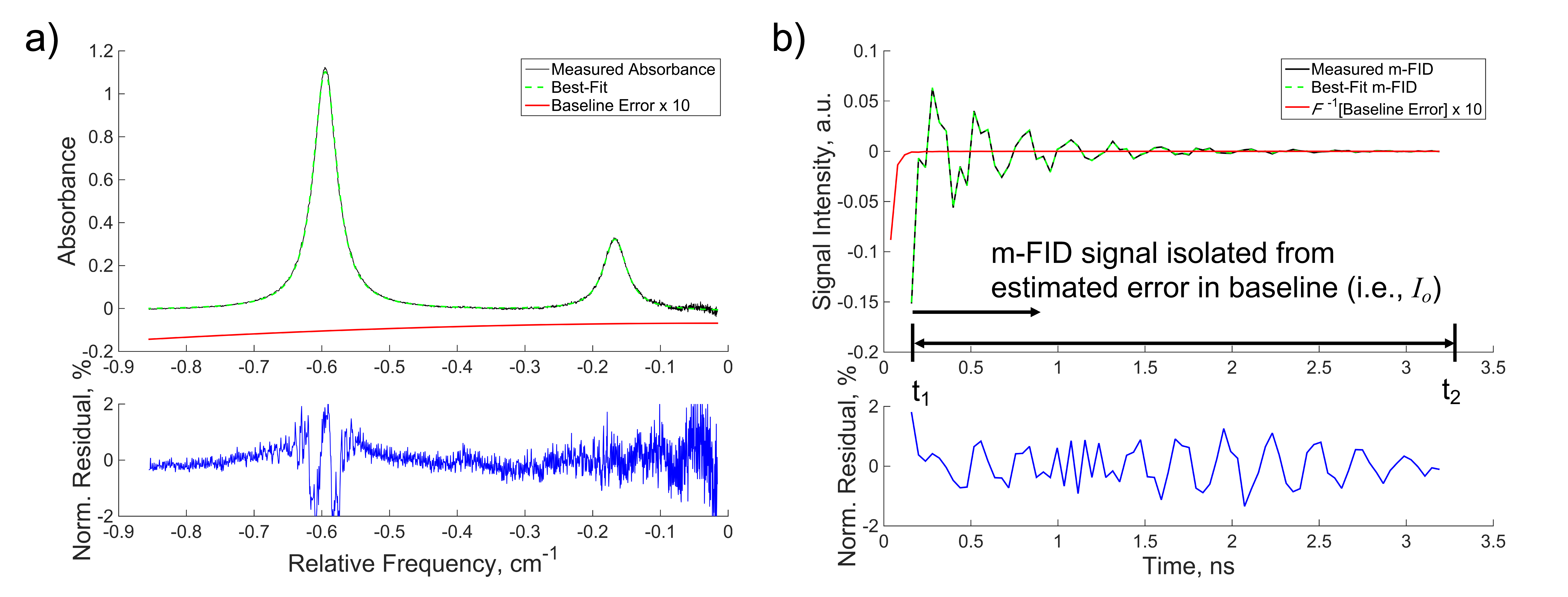}
\caption{(a) Example single-scan measurements of CO's absorbance spectrum in an ethylene-air flame, corresponding best-fit absorbance spectrum, estimated error in $I_o$, and peak-absorbance-normalized residual. (b) Time domain signals corresponding to the spectra shown in (a). The baseline error decays rapidly in the time domain thereby enabling baseline-insensitive measurements of gas properties to be obtained via the m-FID signal despite using a small $t_1$. All spectra and signals shown correspond to a scanned-DA measurement at 1618 K with 11.1\% CO by mole at 1 atm.}
\label{fig: AbsResFID}
\end{figure*}

A non-linear least-squares fitting routine employing the Levenberg-Marquardt algorithm (Matlab's nlinfit) was used to determine the best-fit m-FID signal and corresponding gas conditions. The algorithm seeks to minimize the sum-of-squared error between the measured and simulated m-FID signals at times between $t_1$ and $t_2$ which must be chosen appropriately (discussed later) to isolate the best-fit m-FID signal from error introduced by uncertainty in the estimated $I_o$. In this work, 5 free parameters were employed to adjust the semi-empirical m-FID signal and, more specifically, the underlying absorbance spectrum it corresponds to. The following model inputs were treated as free parameters: (1) the gas temperature, (2) the mole fraction of the absorbing species, (3) a scaling parameter for $\Delta\nu_c$ of all absorption lines, and (4-5) two frequency shift parameters to accurately position the linecenter frequency of each of the dominant absorption lines. Given the large disparity in magnitude between the various free parameters, the temperature was scaled by 10$^{-4}$ and the scaling factor on $\Delta\nu_c$ was scaled by 10$^{-2}$ prior to feeding these inputs to the absorption spectroscopy model called by nlinfit. This scaling is reversed within the absorption spectroscopy model. The gas pressure and optical path length were held constant at the known values. Utilizing a single scaling factor on the collisional widths (e.g., to account for unknown collisional broadening in combustion gas) of all absorption lines is justified here since collisional broadening coefficients for CO's P(0,20) and P(1,14) transitions are similar (e.g., differing by only $\approx$2\% percent for air broadening at the temperatures of interest here). This approach may not be well suited, for example, for measurements of H$_2$O absorbance spectra where collisional-broadening coefficients can vary dramatically between states and collision partners \cite{Goldenstein2015,Rothman2013}. 

\subsection{Selection of the fitting window start time, $t_1$}

Selecting an appropriate value for $t_1$ is critical to maximizing the accuracy of the best-fit parameters (i.e., gas conditions) since this parameter governs which of the strong early-time m-FID signal components are ignored by the least-squares fitting routine. We will show that errors in the simulated m-FID signal which are introduced by errors in the estimated $I_o$ appear at very early times in the m-FID signal, and thus selecting a value of $t_1$ that is too small will retain the influence of those errors. On the other hand, using a value of $t_1$ that is too large (e.g., to avoid all dependence on $I_o$) could correspond to ignoring too much of the molecular-absorption response, thereby making it difficult or impossible to accurately infer the underlying absorbance spectrum and gas conditions it corresponds to. The remainder of this section describes how to use an estimated error in $I_o(\nu)$ (obtained from a spectral-fitting routine) to determine an appropriate value for $t_1$. This approach follows from recognizing that $I_o$ can be described by Eq. \ref{eq: Ioerr}:

\begin{equation} 
I_o(\nu) = I_{o,estimate}(\nu)I_{o,error}(\nu)
\label{eq: Ioerr}
\end{equation}

\noindent where $I_o$ is the true incident laser intensity, $I_{o,estimate}$ is an estimate for $I_o$ (e.g., from baseline fitting), and $I_{o,error}$ is an unknown frequency-dependent correction factor which accounts for the error in $I_{o,estimate}$. In this case, the m-FID signal is given by Eq. \ref{eq: Aterror}:

\begin{multline} 
A(t) = \mathcal{F}^{-1}[\alpha(\nu)] +  \mathcal{F}^{-1}[-ln(I_{o,estimate}(\nu))] \\ + \mathcal{F}^{-1}[-ln(I_{o,error}(\nu))]
\label{eq: Aterror}
\end{multline}

\noindent which shows that the m-FID signal consists of three distinct components with an additive relationship. As a result, to achieve an accurate measurement from least-squares fitting a simulated m-FID signal to a measured m-FID signal, $t_1$ must simply be chosen such that the contribution from $\mathcal{F}^{-1}[-ln(I_{o,error}(\nu))]$ has decayed to zero. It should be noted that the error in $I_{o,estimate}$ can also be accounted for inside the exponential of Beer's law via a frequency-dependent shift in the absorbance. In this case the additive relationship of the three m-FID signal components holds, but the time-domain signal associated with baseline error would be given by $\mathcal{F}^{-1}[-I_{o,error}(\nu)]$ (i.e., differing only by $ln$). This approach was taken here for convenience.

Figure \ref{fig: AbsResFID}a shows an example of a single-scan measurement of CO's absorbance spectrum in the ethylene-air flame, the corresponding best-fit spectrum, the estimated error in $I_o$, and the residual between the measured and best-fit spectrum. The measured absorbance spectrum was calculated using an $I_o$ that was obtained using the traditional method of fitting of a 3rd-order polynomial baseline to the non-absorbing regions of $I_t$. The best-fit spectrum was calculated using a spectral-fitting routine analogous to that described previously for determining the best-fit m-FID signal; however, in addition a 3rd-order polynomial was superimposed onto the simulated absorbance spectrum in an effort to account for and estimate errors in $I_o(\nu)$ that were induced by the imperfect nature of inferring $I_o$ from the traditional method of fitting a polynomial to the "non-absorbing" regions of $I_t$. This approach is later referred to as "Method 2." The coefficients of the polynomial were treated as free-parameters in the model, thereby leading to a total of 9 free parameters (compared to 5 needed for reliable measurements via m-FID signals). Using this method, peak-absorbance-normalized residuals typically <2\% were achieved. In this case, the baseline error (inferred from the polynomial incorporated within the spectroscopic model) varied monotonically from absorbance equivalent values of -0.014 to -0.007. Figure \ref{fig: AbsResFID}b illustrates that the baseline error decays to zero rapidly in the time domain, which is expected given that it varies slowly and smoothly in frequency space. The time required for the estimated baseline error to decay to within 1\% of its initial value was used to determine $t_1$. In this case, $t_1$=0.15 ns and this corresponds to the 4th data point in the time history of the m-FID signal. As a result, the first 3 data points in the time history were ignored by the least-squares fitting routine used to determine the best-fit m-FID signal.

Figure \ref{fig: AbsResFID}b also shows the measured and best-fit m-FID signals at times between $t_1$ and $t_2$. The measured and simulated semi-empirical m-FID signals at times less than $t_1$ and greater than $t_2$ were not used in any manner to determine the best-fit m-FID signal and corresponding gas conditions. The best-fit m-FID signal agrees within 2\% of the measured m-FID signal at all times. The value of $t_2$ (3.25 ns) was chosen to be sufficiently large such that the molecular-absorption response had decayed to within 0.1\% of its initial value, thereby retaining the vast majority of information pertaining to the absorbance spectrum. Using a larger $t_2$ was not found to significantly impact the gas conditions corresponding to the best-fit signal. It should be noted that utilizing both sides (i.e., at the beginning $t_1$ to $t_2$ and end $t_{end}-t_2$ to $t_{end}-t_1$) of the m-FID signal time-history (as done by Cole et al. \cite{Cole2019a}) which approximately mirror each other was not found to significantly (i.e., >0.2\% change) impact the gas conditions corresponding to the best-fit m-FID signal. This may not always be the case depending on the spectrum of the noise in the data.



\begin{figure*}[!t]
\centering
\includegraphics[width=18cm]{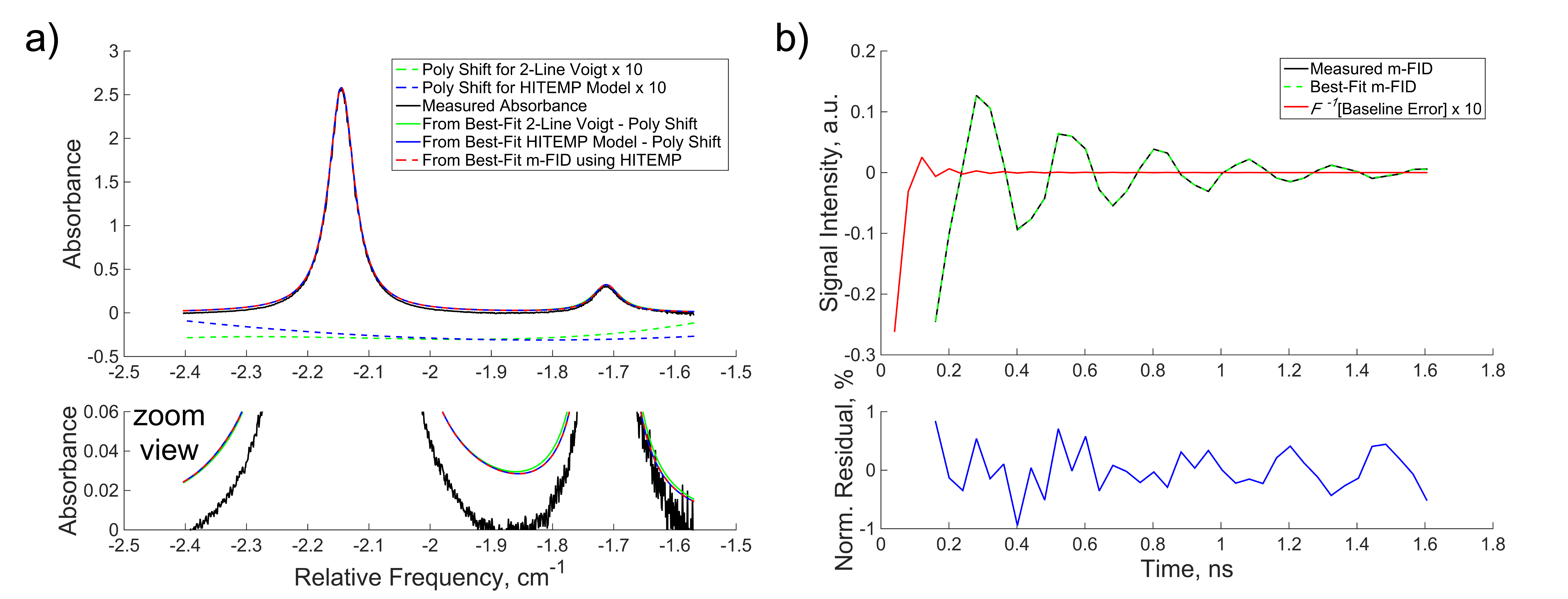}
\caption{(a) Example single-scan measurement of CO absorbance spectra with baseline error and best-fit spectra calculated using Methods 1-3 which address baseline error via a polynomial shift or using the m-FID signal between $t_1$ and $t_2$. (b) Measured and best-fit m-FID signal (calculated using Method 3) corresponding to the measured absorbance spectra shown in (a). The measurements were acquired in a mixture of 2\% CO in N$_2$ at 1 atm and 1034 K.}
\label{fig: mFID_GasCell_Fits}
\end{figure*}

\section{Experimental Results} \label{Sect: ExpResults}

This section presents measurements of gas temperature and CO concentration acquired in a static-gas cell and ethylene-air diffusion flame. Results are presented using the new m-FID approach as well as two traditional data-processing techniques in order to provide proper context for the results obtained using the new m-FID-based signal-processing technique. Specifically, results from the following methods are presented. "Method 1" corresponds to least-squares fitting an absorbance spectrum provided by a two-line Voigt model \cite{Goldenstein2013e} with a 3rd-order polynomial baseline correction, "Method 2" corresponds to least-squares fitting a complete absorbance spectrum (obtained using HITEMP2010 \cite{Rothman2010}) with a 3rd-order polynomial baseline correction, and "Method 3" corresponds to least-squares fitting a simulated, semi-empirical m-FID signal to the measured m-FID signal as described in Section \ref{Sect: FitRoutine}. All methods rely on the same initial estimate for $I_o$ which is obtained from least-squares fitting a 3rd-order polynomial to the non-absorbing regions of $I_t$. Method 1 employs 10 free-parameters ($\nu_o$, $\nu_c$, and $A_{\alpha}$ (the integrated absorbance) for both lines and 4 polynomial coefficients for correcting the baseline) with the Doppler full-width at half-max ($\Delta\nu_D$) fixed according to the temperature obtained from the two-color ratio of integrated absorbance. Method 3 employed 5 free-parameters to model the absorbance spectrum and, hence, m-FID signal (as described in Section \ref{Sect: FitRoutine}), and Method 2 employed an additional 4 free-parameters (for a total of 9) for modeling the baseline error with a 3rd-order polynomial (as done in Method 1).  

Using Method 1, the gas temperature was calculated from the two-color ratio of integrated absorbances provided by the fitting routine, and the mole fraction of CO was calculated from the integrated absorbance of the P(0,20) line. Using Methods 2 and 3, the gas temperature and mole fraction of CO are free-parameters and, therefore, are direct outputs of the fitting routine. In flame experiments, the path length through the flame was assumed to be 1.25 cm (estimated from visible images). The spectroscopic parameters employed by all three methods were taken from HITEMP2010 \cite{Rothman2010} which is known to be accurate for these transitions \cite{Spearrin2014c} and the measurements reported here support this further.

\subsection{Gas-Cell Measurements}

\begin{table*}[!t]
\centering
\caption{Comparison of results obtained using various data-processing techniques for 100 measurements of temperature and CO mole fraction acquired in a static-gas cell at 827 and 1034 K.}
\begin{tabular}{*9c}
\toprule
Method &  \multicolumn{2}{c}{Average} & \multicolumn{2}{c}{Spread} & \multicolumn{2}{c}{Error} & \multicolumn{2}{c}{1-$\sigma$}\\
\midrule
{}                                 & T, K              & X$_{CO}$          & T, K          & X$_{CO}$          & T, \%         & X$_{CO}$,\%   & T, \%         & X$_{CO}$, \%\\
1. 2-Line Voigt + poly             &  916.3, 1077.8    & 0.0224, 0.0216    & 160.7, 75.1   & 3.8E-3, 1.8E-3    & 10.8, 4.2     & 12.0, 8.0     & 10.4, 3.9     & 10.4,	4.9  \\
2. $\alpha$ from HITEMP + poly     &  833.1, 1035.3    & 0.0203, 0.0205    & 24.1, 19.6    & 3.4E-4, 3.3E-4    & 0.7, 0.1      & 1.5, 2.5      & 1.7,	1.1     & 1.0,	1.1  \\
3. m-FID from HITEMP               &  831.6, 1034.5    & 0.0202, 0.0204    & 13.6, 11.7    & 1.1E-5, 2.4E-5    & 0.6, 0.04     & 1.0, 2.0      & 1.1, 0.8      & 0.5,	0.6  \\
\bottomrule
\label{Table: GasCellData}
\end{tabular}
\end{table*}

%

Measurements of temperature and CO mole fraction were acquired at 1 kHz (due to using up-scan and down-scan measurements) over a 100 ms period (100 measurements) in a mixture of 2\% CO in N$_2$ at 1 atm and temperatures of 827 and 1034 K. Figure \ref{fig: mFID_GasCell_Fits}a shows an example of a single-scan measurement of CO's P(0,20) and P(1,14) absorption transitions at 1 atm and 1034 K,  as well as the best-fit spectra corresponding to Methods 1 through 3 and, if applicable, the error in baseline intensity inferred from the polynomial baseline correction (i.e., "poly shift") which was superimposed on the simulated absorbance spectra within the spectral-fitting routine (for Methods 1 and 2 only). It is important to note, that each method was applied to the same spectra with the same initial estimate for the baseline light intensity. As a result, Methods 1, 2, and 3 were all applied to a measurement with the same initial baseline error.  First, the results shown in Figure \ref{fig: mFID_GasCell_Fits}a illustrate that there is an error in $I_o$ of $\approx$1.5-2.5\% depending on frequency and which model is employed, thereby illustrating how the inferred errors in $I_o$ are coupled to the spectroscopic model. Methods 1 and 2 attempt to account for this error via a 3rd-order polynomial (i.e., the "poly shift", best-fit is shown) and Method 3 escapes this error via use of the m-FID signal with a $t_1$ > 0. The gas temperature and CO mole fraction inferred from Method 1, 2, and 3 for this measurement are 1064 K and 2.11\%, 1038 K and 2.05\%, and 1034 K and 2.04\%, respectively. As a result, Methods 2 and 3 provided nearly identical results which agree with expected values within 0-0.4\% for temperature and 2-2.5\% for CO mole fraction. In contrast, the gas temperature and CO mole fraction inferred from Method 1 exhibit a significantly larger error, specifically, 2.9\% for temperature and 5.5\% for CO mole fraction. The best-fit spectra associated with Methods 2 and 3 are virtually identical, as expected given the nearly identical gas conditions associated with each. However, the best-fit spectrum associated with Method 1 exhibits subtle but significant differences (see zoom view within Figure \ref{fig: mFID_GasCell_Fits}a). In addition, the baseline error inferred from Methods 1 and 2 differ significantly, thereby illustrating how it is difficult to reliably infer the error in $I_o$ and, therefore, motivating the use of Method 3 (i.e., the m-FID-based approach presented in Section \ref{Sect: FitRoutine}). For this measurement, it seems that the additional flexibility provided by floating the collisional FWHM of both lines in Method 1 (the 2-line Voigt method) prevented the spectral-fitting routine from accurately inferring the error in the baseline, thereby introducing biases in the integrated absorbance inferred for one or both of the transitions and ultimately leading to errors in the gas conditions corresponding to the best-fit spectrum. This is supported by the fact that the best-fit collisional FWHM for the P(0,20) and P(1,14) lines according to Method 1 were 0.0489 and 0.0537 cm$^{-1}$. This corresponds to a difference of 9.8\% where calculations performed using air-broadening coefficients and temperature exponents from HITEMP2010 \cite{Rothman2010} suggest, albeit assuming an air bath gas, that the collisional FWHM for these lines should agree within 2.2\% at 1034 K.

\begin{figure*}[!t]
\centering
\includegraphics[width=18cm]{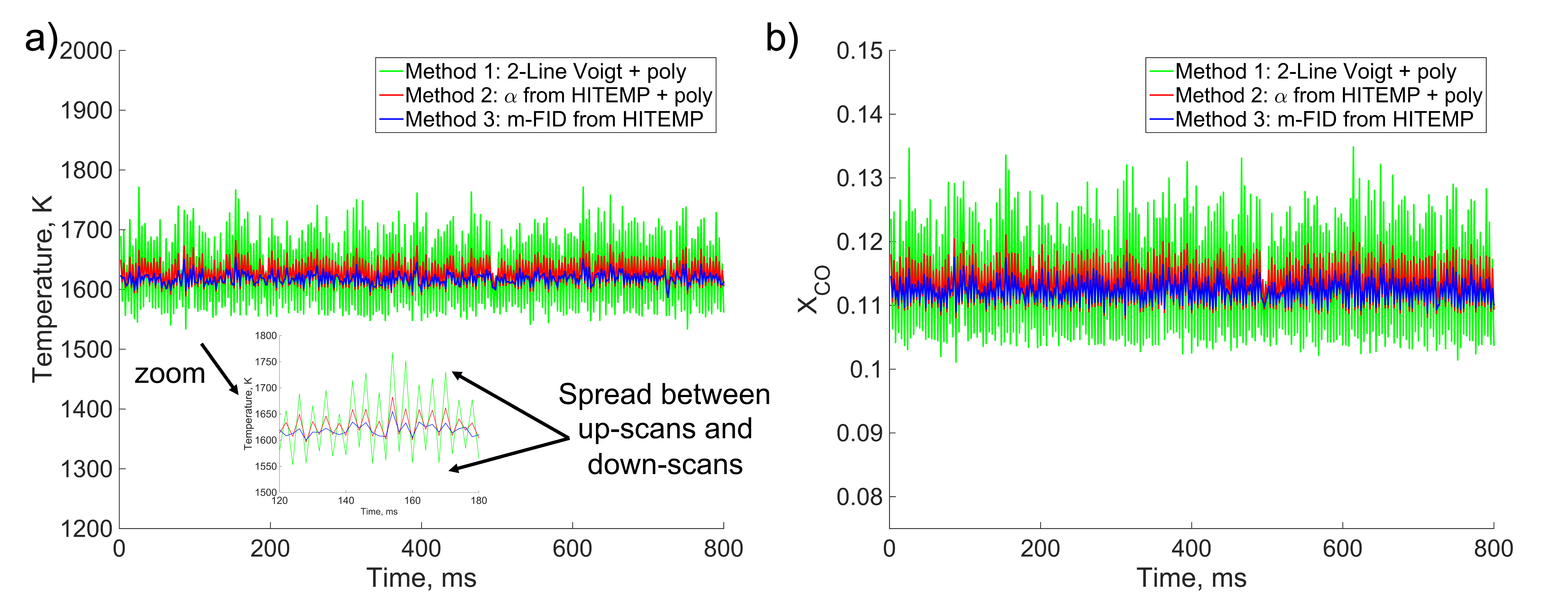}
\caption{Measured time histories of (a) temperature and (b) CO mole fraction acquired using Methods 1-3 for data acquired in an ethylene-air diffusion flame at 1 atm. The results demonstrate that Method 3 provides the most precise measurements.}
\label{fig: Flame}
\end{figure*}

\begin{table*}[!t]
\centering
\caption{Comparison of results obtained using various data-processing techniques for measurements of temperature and CO concentration acquired in a laminar ethylene-air diffusion flame at 1 atm.}
\begin{tabular}{*7c}
\toprule
Method &  \multicolumn{2}{c}{Average} & \multicolumn{2}{c}{Spread} & \multicolumn{2}{c}{1-$\sigma$}\\
\midrule
{}                                  & T, K       & X$_{CO}$  & T, K      & X$_{CO}$     & T, \%       & X$_{CO}$, \%\\
1. 2-Line Voigt + poly             &  1633.3     & 0.115     & 121       & 0.018        & 4.1        & 8.5  \\
2. $\alpha$ from HITEMP + poly     &  1627.1     & 0.113     & 42.8      & 0.008        & 1.4        & 3.5  \\
3. m-FID from HITEMP               &  1617.3     & 0.112     & 12.1      & 0.004        & 0.7        & 1.9  \\
\bottomrule
\label{Table: FlameData}
\end{tabular}
\end{table*}

Table \ref{Table: GasCellData} shows the average value, average spread (i.e., difference) between up-scan and down-scan measurements, error in the average value, and the 1-$\sigma$ precision (i.e., 1 standard deviation) for each dataset which consists of 100 individual measurements (50 up-scans and 50 down-scans). The results illustrate two key findings. First, in all cases, the m-FID based approach (Method 3) was the most accurate. The temperature and CO mole fraction inferred from the best-fit m-FID signal were accurate to within 0.6\% and 1.0\% at 827 K and 0.04\% and 2.0\% at 1034 K. Method 2 provided measurements with slightly larger errors, and Method 1 was considerably less accurate presumably due to its increased sensitivity to the baseline error encountered in this experiment. Second, in all cases, Method 3 provided a smaller measurement precision and smaller spread between up-scan measurements and down-scan measurements (this was also the case in flame experiments, see zoom view within Figure \ref{fig: Flame}). The spread is caused by differences in baseline error between up-scans and down-scans. The influence of this error upon the temperature and CO mole fraction inferred from the data is reduced in Method 3 because the m-FID approach is insensitive to baseline errors. For example, for the dataset acquired at 1034 K, the spread between the mean gas temperature inferred from up-scan measurements and down-scan measurements differed by only 11.7 K (1.1\%) using Method 3, but 19.6 K (1.9\%) for Method 2 and 75.1 K (7.3\%) for Method 1. Method 3 exhibited a spread in temperature measurements that was $\approx$1.7 times smaller than Method 2 and 6.4 to 11.1 times smaller than Method 1. Method 3 also demonstrated similar improvements (factor of 1.37 to 10 for temperature and 1.8 to 20 for CO mole fraction) in the 1-$\sigma$ measurement precision for temperature and CO mole fraction measurements. Collectively these results demonstrate the ability of Method 3 to considerably improve the accuracy and precision of temperature and concentration measurements acquired using LAS with injection-current-tuned lasers.

\subsection{Flame Measurements}
Measurements of temperature and CO mole fraction were acquired in an ethylene-air diffusion flame at 1 atm to evaluate the performance of Method 3 in a test environment with beamsteering, which can introduce time-varying errors in the baseline. Figure \ref{fig: Flame} illustrates measured time histories of temperature and CO mole fraction acquired using Methods 1 through 3 and Table \ref{Table: FlameData} shows the average temperature and CO mole fraction, as well as the spread between up-scans and down-scans and 1-$\sigma$ precision for the 800 ms time histories shown (800 total measurements). The measured time histories illustrate that the flame conditions were quasi-steady during the test. The mean gas temperature and CO mole fraction agree within precision for all three methods. Method 3 provided a significantly smaller spread and precision for temperature and CO mole fraction. Specifically, the spread in temperature provided by Method 3 was 3.5 and 10 times smaller than Methods 2 and 3, respectively. Similarly, the spread in CO mole fraction provided by Method 3 was 2 and 4.5 times smaller compared to Methods 2 and 3, respectively. These results further demonstrate that Method 3 provides superior measurement precision compared to Methods 1 and 2, which further suggests that utilizing the best-fit m-FID signal to determine gas conditions is a more robust technique when errors in the baseline light intensity are present. 

\section{Conclusions}

This manuscript presented a new method which improves the accuracy and precision of LAS measurements of gas properties by isolating the molecular-absorption signal from errors induced by unknown variations in the baseline light intensity. This method relies on least-squares fitting a simulated m-FID signal to a measured m-FID signal in the time domain, where the former is obtained from an estimated $I_o$ and simulated absorbance spectrum. This modified approach is required (in comparison to other m-FID-based techniques \cite{Cole2019a}) when the laser-intensity response and molecular-absorption response do not separate well in the time domain (as is the case in many applications employing TDLs, QCLs, and other light sources which exhibit pronounced intensity tuning). While this particular implementation is not "baseline-free," it was demonstrated that error induced by the estimated $I_o$ can be avoided by ignoring the beginning of the m-FID signal time history in the fitting routine. 

This approach was demonstrated using scanned-wavelength direct-absorption-spectroscopy measurements of CO's P(0,20) and P(1,14) absorption lines using a DFB QCL. Measurements of gas temperature and CO were obtained in a static-gas cell and ethylene-air diffusion flames. The new m-FID-based method demonstrated the ability to provide improved measurement accuracy and precision in all cases compared to two established methods which rely on inferring baseline errors via polynomial corrections. 

The theory and results presented here suggest that this m-FID-based approach can improve the measurement accuracy and precision of a wide range of absorption-spectroscopy diagnostics. 

\section{Acknowledgements}
This work was supported by Grant FA9300-19-P-1506 with Dr. John W. Bennewitz of the Air Force Research Laboratory (AFRL) as program monitor. G.C.M was supported by the National Science Foundation Graduate Research Fellowship Program (NSF GRFP, Grant: 1842166-DGE). In addition, G.B.R. and A.S.M. acknowledge support from the Air Force Office of Scientific Research (AFOSR, Grant: FA9550-17-1-0224) with Dr. Chiping Li as program monitor, and R.K.C. was supported by the National Aeronautics and Space Administration under the Earth and Space Sciences Fellowship program (PLANET18R-0018).


\bibliography{library_forMFID}







\end{document}